\documentclass[conference]{IEEEtran}
\usepackage{fancyhdr}

\usepackage{cite}
\usepackage{amsmath,amssymb,amsfonts,braket}
\usepackage{algorithmic}
\usepackage{graphicx}
\usepackage{textcomp}
\usepackage{xcolor}
\usepackage{tabularx}
\usepackage{booktabs}
\usepackage{balance}
\usepackage{tikz}
\usetikzlibrary{arrows.meta,decorations.pathreplacing,patterns}

\definecolor{lmured}{cmyk}{0.05,1,1,0}
\definecolor{lmublue}{cmyk}{1,0.75,0,0.20}

\graphicspath{{./img/}}

\def\BibTeX{{\rm B\kern-.05em{\sc i\kern-.025em b}\kern-.08em
        T\kern-.1667em\lower.7ex\hbox{E}\kern-.125emX}}

\begin{document}
    \bstctlcite{IEEEexample:BSTcontrol}
    \title{PauLIB: HPC Strategies for Pauli-Arithmetic-Based\\Quantum Computing Workloads and Simulations}

    \author{
    	\IEEEauthorblockN{Florian Krötz}
    	\IEEEauthorblockA{\textit{MNM-Team}, \textit{LMU Munich}, Munich, Germany, florian.kroetz@nm.ifi.lmu.de}
    }

    \maketitle
    \thispagestyle{fancy}
    \fancyhf{}
    \lfoot{\parbox[b]{\textwidth}{\scriptsize
    \copyright~2026 IEEE. Personal use of this material is permitted.
    Permission from IEEE must be obtained for all other uses, in any current or
    future media, including reprinting/republishing this material for
    advertising or promotional purposes, creating new collective works, for
    resale or redistribution to servers or lists, or reuse of any copyrighted
    component of this work in other works.}}
    \renewcommand{\headrulewidth}{0pt}
    \renewcommand{\footrulewidth}{0pt}
	\begin{abstract}
	    As quantum devices become accelerators within classical HPC deployments, the efficiency of the classical partition becomes critical.
Workloads such as Hamiltonian manipulation, observable backpropagation, and Pauli-based compilation reduce to bulk arithmetic on Pauli strings and sums.

We present HPC strategies for this arithmetic, demonstrated in PauLIB, a header-only C++20 library: a bit-packed binary symplectic representation reducing Pauli multiplication to a bitwise XOR and population count; a sorted, struct-of-arrays layout replacing hash maps and exposing contiguous word arrays to SIMD units and GPU warps; and deterministic OpenMP parallelism with race-free output partitioning.

These strategies deliver orders-of-magnitude higher throughput and smaller memory footprint than Qiskit and PauliEngine, plus a speedup from a CUDA outer-product kernel on an H100.
Two case studies, g-sim and truncated Pauli propagation, inherit these gains and scale unchanged from a desktop Arm CPU to a 72-core HPC Arm server.
PauLIB is open source with C++ and Python interfaces.

	\end{abstract}



\section{Introduction}
\label{sec:intro}

The Pauli group is a foundational mathematical structure in quantum information theory~\cite{NielsonChuang, gottesman1997stabilizercodesquantumerror}.
Many quantum computing workflows depend on processing large numbers of Pauli strings, tensor products of single-qubit Pauli operators, across several domains: in quantum chemistry, fermionic Hamiltonians are mapped to Pauli sums via Jordan-Wigner or Bravyi-Kitaev transforms~\cite{OpenFermion};
in quantum error correction, stabilizer codes are maintained as Pauli tableaux~\cite{Stim,aaronson04};
in circuit compilation, Pauli-based computing~\cite{pbc} commutes and reorders non-Clifford rotations expressed as Pauli strings;
and in classical simulation, Pauli propagation algorithms~\cite{PauliPropagation1, PauliPropagation2} and lie-algebraic simulation~\cite{gsim} track how observables evolve through a circuit or under a Hamiltonian without ever representing the exponentially large quantum state.

The computational load grows quickly with problem size.
Molecular Hamiltonians of practical interest can contain millions of Pauli terms~\cite{OpenFermion}.
Each iteration of the Variational Quantum Eigensolver (VQE) requires grouping these terms into mutually commuting fragments and propagating observables through parametrized circuits.
In Pauli propagation, a single non-Clifford rotation can double the number of tracked terms, so simulations routinely execute millions of multiplications and commutation checks per step.
At this scale the throughput of the Pauli algebra back-end becomes the dominant runtime cost, and, crucially, the \emph{same} small set of kernels (string multiplication, commutation checking, duplicate merging, term truncation) dominates all of the workloads above.
This observation is the premise of this paper: optimize the shared kernels once, at the level of data layout and instruction selection, and every workload built on top inherits the speedup.

Existing quantum computing frameworks are not designed for this throughput.
General-purpose tools such as Cirq~\cite{Cirq} and PennyLane~\cite{PennyLane} represent Pauli strings as Python objects or plain character strings, incurring interpreter overhead on every arithmetic operation.
Qiskit~\cite{Qiskit} uses NumPy boolean arrays, which avoids some overhead but stores two full arrays of shape $(M, 2n)$ per Pauli sum, leading to high memory consumption and limited vectorization.
Specialized simulators such as Stim~\cite{Stim} adopt the binary symplectic representation for speed but target the stabilizer formalism only.
The Julia packages PauliPropagation.jl~\cite{PauliPropagation2} and PauliStrings.jl~\cite{paulistringsjl} support general Pauli sums but store them in hash maps (dictionaries), which prevents SIMD vectorization, complicates multi-threaded merging, and, in the case of PauliStrings.jl, leaves the implementation single-threaded.
Further specialized libraries accelerate individual pieces of the problem: state application in fast\_pauli~\cite{fastpauli}, NumPy-array manipulation in PauliArray~\cite{pauliarray}, and measurement partitioning in PSFAM~\cite{psfam}.
PauliEngine~\cite{muller_pauliengine_2026} is the closest in spirit: it is a C++ implementation of the same per-string bitwise arithmetic (Section~\ref{sec:symplectic}), but operates on one Pauli string at a time rather than restructuring a Pauli \emph{sum} for bulk access.
PauLIB's contribution over it is accordingly not a faster single-string kernel, both reduce to the same XOR/popcount trick, but the struct-of-arrays layout (Section~\ref{sec:datastructure}) that keeps thousands of terms cache-contiguous, so the same per-term bit-tricks vectorize and parallelize across a whole sum instead of paying one function call and one potential cache miss per string (Section~\ref{sec:eval} quantifies the resulting gap).
None of these tools provides a unified, SIMD-parallel, multi-threaded back-end for the full generalized Pauli algebra together with the workload layer built on it.
None of these tools addresses how their arithmetic fits into the classical partition of a Q-HPC deployment, where quantum devices increasingly run as accelerators behind classical HPC resource managers~\cite{beck_integrating_2024}.
The simulation and propagation workloads studied here are precisely the classical jobs such orchestration schedules alongside quantum work, and the kernel-level throughput of Sections~\ref{sec:eval}--\ref{sec:workloads} is what determines how efficiently that classical partition is used.

This paper makes two contributions.
\begin{enumerate}
    \item \emph{HPC strategies for Pauli arithmetic.}
    We describe a set of data-structure and parallelization strategies, bit-packed binary symplectic encoding, sorted arrays instead of hash maps, a struct-of-arrays (SoA) transposed layout, two-layer SIMD vectorization, deterministic race-free OpenMP partitioning, and a coalesced GPU kernel mapping, and explain \emph{why} the same bit-packed key serves SIMD units and GPU warps equally well.
    \item \emph{Core-operation benchmarks.}
    We quantify each strategy against the strongest available baselines (PauliEngine, Qiskit) on Hamiltonian outer products, memory footprint, and GPU offload.
\end{enumerate}
All strategies are implemented in PauLIB~\cite{paulib_github}, an open-source, header-only C++20 library with Python bindings.
We demonstrate both contributions end to end with two complete simulation workloads built on the same kernels, Lie-algebraic simulation (g-sim)~\cite{gsim} and truncated Pauli propagation~\cite{PauliPropagation1}, validated against dense linear-algebra references, showing that the kernel-level speedups of Section~\ref{sec:eval} carry through to workload level (Section~\ref{sec:workloads}).

The rest of this paper is organized as follows.
Section~\ref{sec:pauli} reviews Pauli strings, Pauli sums, and the core operations on them.
Section~\ref{sec:datastructure} presents the Pauli-sum data structure and its suitability for SIMD and GPU execution.
Section~\ref{sec:eval} reports core-operation benchmarks against other software.
Section~\ref{sec:workloads} presents the two workload case studies with benchmarks.
Section~\ref{sec:conclusion} concludes.


\section{Pauli Strings, Sums, and Core Operations}
\label{sec:pauli}

\subsection{Pauli Operators and Strings}

The four single-qubit \emph{Pauli operators} form a basis for all single-qubit observables and unitaries:
\begin{align*}
	I &= \begin{bmatrix} 1 & 0 \\ 0 & 1 \end{bmatrix}, &
	Z &= \begin{bmatrix} 1 &  0 \\ 0 & -1 \end{bmatrix}, \\[4pt]
	X &= \begin{bmatrix} 0 &  1 \\ 1 &  0 \end{bmatrix}, &
	Y &= \begin{bmatrix} 0 & -i \\ i &  0 \end{bmatrix}.
\end{align*}
Physically, $Z$ is a phase flip ($\ket{1} \mapsto -\ket{1}$), $X$ is a bit flip ($\ket{0} \leftrightarrow \ket{1}$), and $Y = iXZ$ combines both effects with an additional phase factor.
Together with phase factors $\{\pm1, \pm i\}$, they generate the single-qubit Pauli group.
Each Pauli matrix is its own inverse ($P^2 = I$) and Hermitian ($P^\dagger = P$).
Distinct non-identity Pauli matrices \emph{anti-commute} ($AB = -BA$), and their products always yield another Pauli operator multiplied by a phase:
\begin{alignat*}{3}
	ZX &=~~iY, & \quad YZ &=~~iX, & \quad XY &=~~iZ \\
	XZ &= -iY, & \quad ZY &= -iX, & \quad YX &= -iZ.
\end{alignat*}
These nine products are the only arithmetic needed to multiply any two Pauli operators;
the sign of the phase is determined by the order of the operands.

An $n$-qubit \emph{Pauli string} is a tensor product of $n$ single-qubit Pauli operators, one per qubit:
\begin{align}
	P = \bigotimes_{i=1}^{n} P_i, \quad P_i \in \{I, X, Y, Z\}.
	\label{eq:paulistring}
\end{align}
Multiplying two Pauli strings proceeds qubit by qubit, accumulating the per-position phases into a global phase.
Two Pauli strings commute if the number of qubit positions at which their factors anti-commute is \emph{even};
they anti-commute if that number is \emph{odd}.
This parity test is the only computation needed to decide commutativity, and it maps directly to a bitwise inner product over the binary representation introduced below.

A \emph{Pauli sum} is a linear combination of Pauli strings:
\begin{align}
	H = \sum_{j} c_j P_j, \quad c_j \in \mathbb{C}.
	\label{eq:paulisum}
\end{align}
When all $c_j$ are real, $H$ is Hermitian and represents a quantum observable.
The central example in quantum chemistry is the molecular Hamiltonian, which can contain millions of terms for molecules of practical interest~\cite{OpenFermion}.

\subsection{Conjugation and Pauli Rotations}
\label{sec:rotations}

Many algorithms do not track the quantum state directly.
Instead, they track how Pauli strings, observables or stabilizers, transform as gates are applied: a state on $n$ qubits requires $2^n$ amplitudes, while a Pauli string requires only $2n$ bits.
The transformation of an observable $Q$ under a gate $U$ is the \emph{conjugation}
\begin{align}
	Q' = U^\dagger\, Q\, U.
	\label{eq:conjugation}
\end{align}
Every unitary gate can be expressed as a product of \emph{Pauli rotations}
\begin{align}
	U_P(\theta) = e^{-i\frac{\theta}{2}P}
	            = \cos\!\left(\tfrac{\theta}{2}\right)I
	              - i\sin\!\left(\tfrac{\theta}{2}\right)P,
	\label{eq:paulirot}
\end{align}
where $P$ is a Pauli string and $\theta \in \mathbb{R}$.
If $Q$ commutes with $P$, conjugation leaves $Q$ unchanged.
If $Q$ anti-commutes with $P$, three cases arise.

\paragraph{Pauli gate ($\theta = \pi$)}
$Q' = PQP = -Q$: the observable flips sign, and the sum keeps its size.

\paragraph{Clifford gate ($\theta = \pi/2$)}
$Q' = iPQ$: the observable becomes a different, but still single, Pauli string.
Clifford circuits therefore map Pauli strings to Pauli strings~\cite{silva_clifford_2025} and never grow a Pauli sum.
Because conjugation by any unitary is an algebra automorphism, the same relabeling extends from a single Pauli string to an entire rotation: for anticommuting $P, P'$,
\begin{align}
	U_{P'}(\phi)\, U_P\!\left(\tfrac{\pi}{2}\right)
	= U_P\!\left(\tfrac{\pi}{2}\right)\, U_{iPP'}(\phi).
	\label{eq:pbc_commute}
\end{align}
An arbitrary-angle rotation commutes past a fixed Clifford gate by relabeling its Pauli generator alone, with the rotation angle $\phi$ unchanged, the mechanism Pauli-based computing~\cite{pbc, litinski_game_2019} uses to reorder and collect non-Clifford rotations expressed as Pauli strings.

\paragraph{Non-Clifford rotation}
For a general angle,
\begin{align}
	Q' = \cos(\theta)\,Q + i\sin(\theta)\,PQ.
	\label{eq:nonclifford}
\end{align}
One Pauli string becomes a superposition of two, each with a trigonometric coefficient;
applied to a sum of $M$ terms, a single non-Clifford gate can grow the sum to $2M$ terms.
This exponential growth is the central computational challenge of Pauli propagation, controlled in practice by truncating terms with small coefficients or large weight (Section~\ref{sec:workloads}).

\subsection{Binary Symplectic Encoding}
\label{sec:symplectic}

An $n$-qubit Pauli string is mapped to a bit vector $(x \mid z)$, where $x$ and $z$ represent the bit-flip and phase-flip components (Table~\ref{tab:pauli_encoding}).
PauLIB packs each Pauli string into $2n$ bits of operator content plus 2 bits of global phase ($\pm 1, \pm i$), stored in $\lceil n/64 \rceil$ 64-bit words per component.

\begin{table}[h]
	\centering
	\caption{Symplectic Mapping and Bit-Encoding Logic}
	\label{tab:pauli_encoding}
	\begin{tabular}{@{} c cc l @{}}
		\toprule
		\textbf{Pauli Operator} & \textbf{$X$-bit} & \textbf{$Z$-bit} & \textbf{Physical Mapping} \\
		\midrule
		$I$  & 0 & 0 & Identity \\
		$Z$  & 0 & 1 & Phase Flip \\
		$X$  & 1 & 0 & Bit Flip \\
		$Y$  & 1 & 1 & Combined $iXZ$ \\
		\bottomrule
	\end{tabular}
\end{table}

\paragraph{Bit-parallel multiplication}
Because Pauli operators are their own inverses, the operator part of a product $A \cdot B$ is obtained word by word with bitwise XOR:
\begin{align*}
	x[k] &\leftarrow x[k] \oplus b_x[k], &
	z[k] &\leftarrow z[k] \oplus b_z[k],
\end{align*}
for $k \in \{0, \dots, \lceil n/64 \rceil - 1\}$.
The global phase of the product is $i^s$ with $s \in \{0,1,2,3\}$, accumulated from the anti-commuting qubit positions with bitwise logic and popcount instructions following the methodology of Stim~\cite{Stim}, without iterating over individual qubits.

\paragraph{Commutation check}
Two strings with symplectic vectors $v_1, v_2$ commute iff their symplectic inner product
\begin{align*}
	[v_1,v_2] = \bigoplus_{i=1}^{n}
	\left(x_i^{(1)}z_i^{(2)} \oplus z_i^{(1)}x_i^{(2)}\right)
\end{align*}
vanishes.
Each summand detects whether the single-qubit factors at position $i$ anti-commute;
the XOR-accumulation computes the parity of anti-commuting positions.

\subsection{The Core Operation Inventory}
\label{sec:coreops}

The workloads of Section~\ref{sec:workloads} are compositions of a small inventory of operations, all of which reduce to the bitwise kernels above:

\begin{itemize}
	\item \emph{Multiplication} $P_a \cdot P_b$: XOR + popcount (the phase is carried in the result).
	\item \emph{Commutation check}: symplectic inner product, XOR-AND + popcount parity.
	\item \emph{Commutator} $[A,B]$: for Pauli strings, $[P_a, P_b] = 2P_aP_b$ when they anti-commute and $0$ otherwise, one commutation check, then at most one multiplication per pair.
	\item \emph{Conjugation / rotation}~\eqref{eq:nonclifford}: one commutation check and one multiplication per affected term.
	\item \emph{Collate}: after multiplication or propagation, duplicate strings must be merged by summing coefficients (Section~\ref{sec:collate}).
	\item \emph{Truncation}: discard terms by coefficient magnitude, by operator weight (number of non-identity sites, one popcount per term), or keep only the $M$ largest terms.
	\item \emph{Trace pairings}: $\operatorname{tr}(P_aP_b) = 2^n\delta_{ab}$ makes traces, Frobenius norms, and expectation values term-local: $\braket{0^n|P|0^n}$ is $\pm1$ for $I/Z$-only strings and $0$ otherwise.
\end{itemize}

Every operation in this inventory is $O(\lceil n/64 \rceil)$ per term or term pair, independent of the Hilbert-space dimension $2^n$.
The remaining question, and the subject of the next section, is how to organize millions of such terms in memory so that modern hardware can stream through them at full width.


\section{A Pauli-Sum Data Structure for SIMD and GPUs}
\label{sec:datastructure}

The kernel inventory of Section~\ref{sec:coreops} is cheap per term, a handful of bitwise instructions, so at scale the performance question is not arithmetic but \emph{data movement}: how terms are laid out, how duplicates are merged, and whether the inner loops are branch-free and contiguous.
This section presents the two complementary data types used by PauLIB and explains why the same layout decisions serve CPU SIMD units and GPU warps alike.

\subsection{Memory Layout}
\label{sec:layout}

A single \texttt{PauliString} is one fixed-size, packed struct (Fig.~\ref{fig:paulistring_layout}): $W = \lceil n/64 \rceil$ 64-bit $X$-words, $W$ 64-bit $Z$-words, and a flags byte holding the phase exponent $k$ of $i^k$.
Qubit $j$ occupies bit $j \bmod 64$ of word $\lfloor j/64 \rfloor$ in both bit planes, so a qubit lookup is one shift and one mask.
With the flags byte padded to a full 64-bit word, a string occupies exactly $16W + 8$ bytes, needs no heap allocation, and is trivially copyable.
This layout is convenient for single-term operations and direct indexing.

\begin{figure}[!htb]
    \centering
    \resizebox{\linewidth}{!}{%
    \begin{tikzpicture}[
        word/.style={draw, minimum width=10mm, minimum height=6mm, inner sep=0pt, font=\small},
        brc/.style={decorate, decoration={brace, amplitude=4pt}, thick, black!60},
        lbl/.style={font=\footnotesize, above=4pt}]
      \foreach \lab/\xa/\w/\col in {{$x_0$}/0/1.0/lmured!25, {$x_{W-1}$}/1.6/1.0/lmured!25,
                                    {$z_0$}/2.6/1.0/lmublue!25, {$z_{W-1}$}/4.2/1.0/lmublue!25,
                                    {$f$}/5.2/0.5/black!8, {}/5.7/1.3/white}
        \node[word, minimum width=\w cm, fill=\col, anchor=west] at (\xa,0) {\lab};
      \foreach \xa in {1.0,3.6} {
        \draw (\xa,0.3) -- (\xa+0.6,0.3) (\xa,-0.3) -- (\xa+0.6,-0.3);
        \node[font=\small] at (\xa+0.3,0) {$\cdots$};
      }
      \fill[pattern=north east lines, pattern color=black!30] (5.7,-0.3) rectangle (7.0,0.3);
      \draw[brc] (0,0.4) -- node[lbl, text=lmured] {$x$ bits: $W\times 64$\,bit} (2.6,0.4);
      \draw[brc] (2.6,0.4) -- node[lbl, text=lmublue] {$z$ bits: $W\times 64$\,bit} (5.2,0.4);
      \draw[brc] (5.2,0.4) -- node[lbl, text=black!70] {8\,B} (7.0,0.4);
      \node[font=\scriptsize, text=black!60, anchor=north] at (6.35,-0.35) {pad};
      \foreach \b in {0,...,7} {
        \pgfmathsetmacro\xb{5.65 - 0.3*\b}
        \node[draw, minimum width=3mm, minimum height=3.5mm, inner sep=0pt, font=\scriptsize,
              fill=black!8] (b\b) at (\xb,-1.1) {\ifnum\b<2 \else\textcolor{black!40}{0}\fi};
      }
      \node[font=\scriptsize] at (b1) {$s$};
      \node[font=\scriptsize] at (b0) {$i$};
      \draw[black!50] (5.2,-0.3) -- (b7.north west) (5.7,-0.3) -- (b0.north east);
      \node[font=\footnotesize, anchor=east] at (3.3,-1.1) {phase $i^k$, $k = 2s + i$};
      \node[font=\footnotesize, anchor=north west, align=left] at (-0.05,-1.45)
        {qubit $j$ $\mapsto$ bit $j \bmod 64$ of word $\lfloor j/64\rfloor$\\
         e.g.\ $X_0 Y_1 Z_2$: \ $x_0 = \ldots 0011_2$, \ $z_0 = \ldots 0110_2$};
    \end{tikzpicture}}
    \caption{Memory layout of \texttt{PauliString} for $n$ qubits, $W=\lceil n/64\rceil$.
             $W$ 64-bit $X$-words and $W$ 64-bit $Z$-words are followed by one 8-byte word holding the flags byte $f$ (sign bit $s$ and imaginary-phase bit $i$, phase $i^{k}$ with $k = 2s+i$) and alignment padding.}
    \label{fig:paulistring_layout}
\end{figure}

For sums, \texttt{PauliSum} transposes this layout into a \emph{Struct-of-Arrays} (SoA) (Fig.~\ref{fig:paulisum_soa}): word $w$ of the $X$-plane of every term is stored in one array, likewise for the $Z$-plane, the flags, and the coefficients.
A sum with $M$ terms therefore consists of $2W$ contiguous arrays of $M$ 64-bit words, one array of $M$ flags bytes, and one array of $M$ coefficients.
A term is no longer a contiguous object but one row across all arrays.

\begin{figure}
    \centering
    \resizebox{\linewidth}{!}{%
    \begin{tikzpicture}[
        cell/.style={draw, minimum width=8mm, minimum height=5mm, inner sep=0pt},
        head/.style={font=\small, anchor=south},
        idx/.style={font=\footnotesize, anchor=east, text=black!60},
        brc/.style={decorate, decoration={brace, amplitude=4pt}, thick, black!60}]
      \foreach \lab/\xx/\col in {{$x_0$}/0/lmured!25, {$x_{W-1}$}/1.4/lmured!25,
                                 {$z_0$}/2.5/lmublue!25, {$z_{W-1}$}/3.9/lmublue!25,
                                 {$f$}/5.0/black!8, {$c$}/6.1/black!8} {
        \node[head] at (\xx,0.3) {\lab};
        \foreach \r in {0,1} \node[cell, fill=\col] at (\xx,-\r*0.5) {};
        \node[font=\small] at (\xx,-0.95) {$\vdots$};
        \node[cell, fill=\col] at (\xx,-1.5) {};
      }
      \foreach \xx in {0.7,3.2} \node[font=\small] at (\xx,-0.5) {$\cdots$};
      \draw[brc] (-0.4,0.85) -- node[above=4pt, font=\footnotesize, text=lmured]
        {$x$ bits of $P_k$} (1.8,0.85);
      \draw[brc] (2.1,0.85) -- node[above=4pt, font=\footnotesize, text=lmublue]
        {$z$ bits of $P_k$} (4.3,0.85);
      \draw[brc] (4.6,0.85) -- node[above=4pt, font=\footnotesize, text=black!70]
        {phase} (5.4,0.85);
      \draw[brc] (5.7,0.85) -- node[above=4pt, font=\footnotesize, text=black!70]
        {coeff.} (6.5,0.85);
      \node[idx] at (-0.62,0)    {term 1};
      \node[idx, text=black, font=\footnotesize\bfseries] at (-0.62,-0.5) {term $k$};
      \node[idx] at (-0.62,-1.5) {term $M$};
      \draw[dashed, thick, black!70, rounded corners=2pt] (-0.47,-0.27) rectangle (6.6,-0.73);
      \draw[{Stealth}-{Stealth}, black!60] (6.85,0.25) -- (6.85,-1.75);
      \node[rotate=90, anchor=north, font=\footnotesize, align=center, text=black!60]
        at (6.95,-0.75) {contiguous\\in memory};
    \end{tikzpicture}}
    \caption{Memory layout of \texttt{PauliSum}.
             For a sum of $M$ terms with $W=\lceil n/64\rceil$ words per bit plane, each column is one contiguous array of $M$ entries: $W$ arrays of 64-bit $X$-words, $W$ arrays of 64-bit $Z$-words, one array of flags bytes ($f$), and one coefficient array ($c$).
             Term $k$ (dashed) is spread across all columns at row $k$.}
    \label{fig:paulisum_soa}
\end{figure}

The SoA transposition is critical for bulk operations.
When a single-qubit gate is applied to qubit $c$, only the arrays at word index $\lfloor c/64 \rfloor$ are accessed: the compiler sees a simple loop over $M$ independent 64-bit integers, all other data stays out of the cache, and the loop vectorizes automatically.

\subsection{Sorted Arrays Instead of Hash Maps}
\label{sec:collate}

Hash maps offer $O(1)$ average-case insertion and are the representation of choice in \cite{PauliPropagation2,paulistringsjl}.
Their pointer-heavy node structure, however, causes frequent cache misses at scale, defeats SIMD vectorization, and serializes multi-threaded merging.
PauLIB instead keeps Pauli sums in flat vectors and restores canonical form with a \emph{collate} step: an index array is sorted using the bit-packed words as a fixed-width lexicographic key ($O(1)$ per comparison, $O(M \log M)$ total), then a single linear scan accumulates the coefficients of adjacent identical strings.
The merge phase is branch-predictable, achieves near-peak memory bandwidth, and leaves the sum sorted, immediately usable for binary-search lookups and SIMD bulk passes.
Truncation composes naturally with this representation: coefficient cutoffs and weight filters are single compaction passes, and keeping the $M$ largest terms is one \texttt{nth\_element} selection, no rehashing, no node deallocation.

Above a size threshold the collate step itself runs thread-parallel, and it does so \emph{deterministically}: the index array is divided into a \emph{fixed} number of chunks (independent of the thread count), each chunk is stable-sorted concurrently, and the sorted runs are combined with stable pairwise merges;
the duplicate merge is then parallelized over the resulting key segments, summing each segment's coefficients in sorted order.
Because the chunk grid does not depend on \texttt{OMP\_NUM\_THREADS}, the output, including the floating-point summation order of duplicate terms, is bit-identical for every thread count, which keeps parallel runs reproducible and directly comparable.

\subsection{Why This Layout Fits SIMD}
\label{sec:simd}

PauLIB relies on compiler auto-vectorization in two places, both enabled directly by the SoA layout and by OpenMP's SIMD annotations.
Intrinsics or SIMD libraries did not provide a significant speedup.
Modern compilers are excellent at optimizing bit operations.

The first is gate application.
Because all $M$ values for the relevant word occupy one contiguous array, the loop body for gates such as H, S, and CNOT is branch-free and operates on independent elements, so the compiler emits AVX2 or AVX-512 instructions, processing 4 or 8 strings per cycle.

The second is the Pauli-sum outer product $H_1 \cdot H_2$, which requires $|H_1| \cdot |H_2|$ string multiplications: the innermost loop, over the left-hand-side index $j$ for one fixed right-hand-side term, is again vectorized using OpenMP, with the full multiplication kernel, XOR for the bit update, AND plus XOR for the anti-commutation mask, and the bitwise phase accumulators, executing lane-wise across $j$.
The key point is that \emph{vector lanes range over terms, not over qubits}: the SoA transposition turns $M$ independent term operations into unit-stride vector loads, which an array-of-structs layout cannot offer, and it does so on any target ISA the build is configured for, without a separate code path per architecture.

\subsection{Thread-Level Parallelism}
\label{sec:omp}

For the outer product, the result array is pre-allocated to size $|H_1| \cdot |H_2|$ and partitioned into contiguous blocks, one per right-hand-side term.
The outer loop is distributed across cores using OpenMP with static scheduling;
each thread writes to a disjoint output block, so there are no data races and no synchronization during the computation.
The commutator kernel follows the same pattern with thread-local output buffers, as does the rotation-gate split of Pauli propagation, whose per-chunk buffers are concatenated in fixed chunk order for the same thread-count-independent determinism as the parallel merge.
Section~\ref{sec:threading} measures the resulting speedup.

\subsection{Why the Same Layout Fits the GPU}
\label{sec:gpu_design}

The SoA flat layout maps onto CUDA with no reshaping: device buffers are byte-for-byte copies of the host arrays.
The GPU outer-product kernel assigns one thread to each output pair $(j, i)$, $n \times m$ threads in total, with the flat indexing $\texttt{lhs\_x}[k \cdot n + j]$ for word $k$ of LHS term $j$.
Consecutive threads share the same $i$ and have consecutive $j$, so all three memory streams behave optimally: LHS loads and output stores are coalesced (stride-one per thread), and the RHS access is a warp-wide broadcast of a single scalar.
The phase computation gives the GPU a further structural advantage over the CPU: the two symplectic phase accumulators $c_1, c_2$, which on the CPU form per-output-row arrays that spill out of L1 for large sums, live in two private 64-bit \emph{registers} per thread.
The word loop is fully unrolled at compile time, so the entire multiplication executes from registers.
In short, the layout decisions taken for SIMD, transposed contiguous word arrays, fixed-width keys, no pointers, are exactly the decisions that give coalesced, divergence-free GPU execution;
nothing about the data structure had to change for the port.

The commutator, the other pairwise kernel in the core-operation inventory of Section~\ref{sec:coreops}, ports the same way: one thread per pair evaluates the symplectic parity entirely in registers and returns early for the commuting majority;
anticommuting pairs claim an output slot through a global atomic counter and write the product term.
The result is compacted to its exact size on the device, ready for immediate reuse without a host round-trip.


\section{Core-Operation Benchmarks}
\label{sec:eval}

This section quantifies the strategies of Section~\ref{sec:datastructure} on the isolated core operations;
Section~\ref{sec:workloads} then measures complete workloads.

The benchmarks in this paper, core-operation and workload alike, run on an \textbf{NVIDIA GH200} node and one or more \textbf{Apple M4~Pro} machines (Section~\ref{sec:workloads} gives each workload table's specific machine).
This section uses the GH200 node's \textbf{Grace CPU} (Arm Neoverse~V2, 72 cores at 2.87~GHz, 1~MiB L2 per core, 114~MiB shared L3) for CPU benchmarks and its \textbf{H100 SXM} GPU (96~GB HBM3 at $\sim$3.35~TB/s) for GPU benchmarks.
Code was compiled with GCC~13.2.0 using \texttt{-O3 -march=native -ffast-math} (\texttt{-mcpu=native} on Arm);
fast-math affects only the scalar coefficient arithmetic, since all Pauli bit logic is integer.
Multi-threaded runs use one thread per physical core (72 threads).
Each timing is one call's wall-clock time, sampled 15--50 times after warm-up over three processes, with competitors interleaved with PauLIB.
We report medians and interquartile ranges (IQR; error bars) and speedups as ratios of medians with 95\% bootstrap confidence intervals;
no outliers are removed, and raw samples are in the artifact.

PauLIB is compared against PauliEngine~\cite{muller_pauliengine_2026} and Qiskit~\cite{Qiskit}.
For each task we benchmark the strongest publicly available baseline that exposes the operation through a supported API: multiplication against PauliEngine and Qiskit (the two frameworks with first-class dense Pauli-product primitives, with algebraic equivalence of the results verified before timing), and the memory comparison against Qiskit, the only baseline whose representation reaches $10^6$ terms at 500 qubits.

\begin{figure}[t]
	\centering
	\includegraphics[width=\columnwidth]{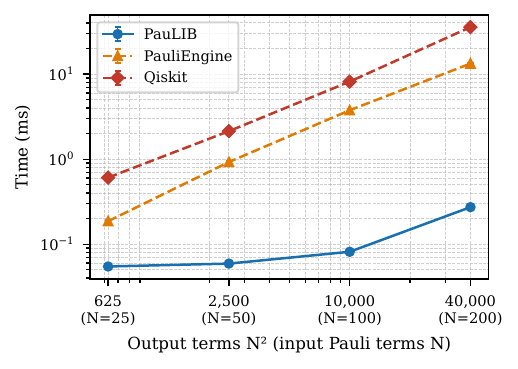}
	\caption{PauliSum$\times$PauliSum outer product at 500 qubits vs.\ output terms $N^2$ (log-log; medians, IQR bars).
		PauLIB (solid) is $3$--$49\times$ below PauliEngine and $11$--$131\times$ below Qiskit (dashed); the gap widens with $N$.}
	\label{fig:sum_mul}
\end{figure}

\subsection{Hamiltonian Outer Product}
\label{sec:sum_mul}

Fig.~\ref{fig:sum_mul} shows the time for the full outer product $H_1 \cdot H_2$ of two random 500-qubit sums of size $N$, producing $N^2$ terms.
PauliEngine and Qiskit scale as $O(N^2)$ throughout, while PauLIB's time is dominated by fixed parallel-launch and allocation overhead ($\approx$60--90~$\mu$s) up to $N = 100$, so the gap widens with $N$.
At $N = 200$, PauLIB completes in 0.27~ms (IQR 0.27--0.28~ms), \textbf{49$\times$} faster than PauliEngine (95\% CI 48--49) and \textbf{131$\times$} faster than Qiskit (129--133);
at $N = 25$ the factors are 3.4$\times$ and 11$\times$.
The gap to PauliEngine specifically is architectural, not algorithmic: both reduce single-string multiplication to the same bit-packed XOR/popcount kernel (Section~\ref{sec:symplectic}), but PauliEngine pays that cost once per string through individual calls, while PauLIB's struct-of-arrays layout (Section~\ref{sec:datastructure}) applies it across an entire sum from contiguous, SIMD-vectorized arrays.

\subsection{Memory Footprint}
\label{sec:memory}

At 500 qubits, PauLIB stores a term in $2 \times 8$ words plus one double, about 137 bytes.
Constructing a sum of $10^4$, $10^5$, and $10^6$ terms increases RSS by \textbf{9}, \textbf{17}, and \textbf{135~MB}, versus \textbf{83}, \textbf{162}, and \textbf{1\,051~MB} for Qiskit's boolean NumPy arrays, a \textbf{7.8--9.5$\times$} reduction across scales (median of five fresh processes per size; PauLIB's RSS is tight across all three sizes, while Qiskit's $10^5$-term point shows a bimodal $\sim$17~MB spread across processes, likely an allocator effect, with the other two sizes within 1~MB).
The reduction has a direct consequence: at 500 qubits and a few million terms, the PauLIB sum stays within a few hundred megabytes, while the Qiskit representation's much steeper growth pushes it into the multi-gigabyte range at the same scale.

\subsection{GPU Outer Product}
\label{sec:gpu}

\begin{figure}[t]
    \centering
    \includegraphics[width=\columnwidth]{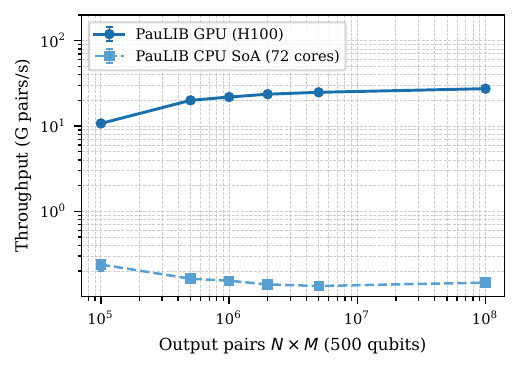}
    \caption{Outer-product throughput at 500 qubits on an NVIDIA GH200 (medians of 150 runs, IQR bars).
             The H100 kernel (solid) reaches 27.4~G terms/s, 187$\times$ above the 72-core Grace CPU kernel, whose flat curve indicates bandwidth saturation.}
    \label{fig:gpu_mul}
\end{figure}

Fig.~\ref{fig:gpu_mul} shows outer-product throughput for the kernel of Section~\ref{sec:gpu_design} on the same GH200 node: the H100 GPU against its own Grace CPU, across problem sizes from $10^5$ to $10^8$ output terms.
The 72-core Grace CPU's curve is flat across that entire range, a signature of a memory-bandwidth-bound kernel, whose phase-accumulator arrays spill out of L1 for large sums.
The GPU kernel, whose accumulators are thread-private registers, has no such ceiling: its throughput keeps climbing with problem size, reaching \textbf{27.4~G terms/s} at $10^8$ terms.
At $n = m = 10\,000$ that gap means the GPU completes the outer product in 3.65~ms versus 684~ms on the CPU, a \textbf{187$\times$} speedup (95\% CI 187--188$\times$).

\subsection{Thread-Level Parallelism}
\label{sec:threading}

\begin{figure}[t]
    \centering
    \includegraphics[width=\columnwidth]{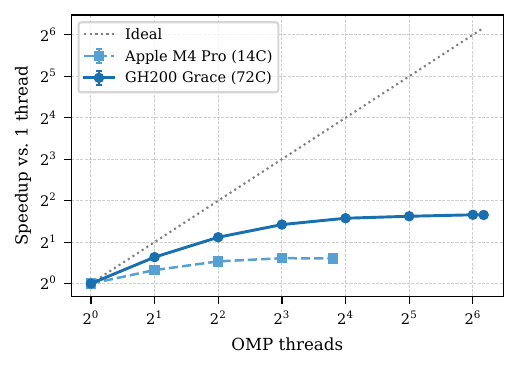}
    \caption{Outer-product speedup vs.\ OMP thread count (500 qubits, $9 \times 10^6$ output pairs; medians of 45 runs, bars: 95\% CI).
             Both platforms saturate well below the ideal (dotted) line, consistent with the memory-bandwidth-bound throughput of Section~\ref{sec:gpu}.}
    \label{fig:thread_scaling}
\end{figure}

Section~\ref{sec:omp} distributes the outer product across cores with one disjoint output block per right-hand-side term, so the parallel region is race-free and synchronization-free by construction.
Fig.~\ref{fig:thread_scaling} measures the resulting speedup: a 500-qubit, $3\,000 \times 3\,000$-term outer product ($9 \times 10^6$ output pairs) timed across the full thread range on both machines above.
Speedup is sub-linear on both, \textbf{2.7$\times$} at 8 threads and \textbf{3.2$\times$} at 72 threads on the GH200 node's Grace CPU, \textbf{1.5$\times$} at 8 threads on the M4~Pro's 14 cores, essentially flat beyond that, because the kernel is memory-bandwidth-bound, not compute-bound: the same characteristic behind the CPU throughput ceiling of Section~\ref{sec:gpu} (123--145~M pairs/s at eight or more threads).
Parallelism still pays for itself well past the point of diminishing per-core returns, since it carries no synchronization overhead;
it does not, however, turn the outer product into an embarrassingly parallel kernel in the compute-bound sense.

\section{Workload Case Studies}
\label{sec:workloads}

The premise of Section~\ref{sec:intro} was that complete simulation workloads are compositions of the core-operation inventory and therefore inherit the kernel-level speedups.
This section substantiates that claim with two workloads implemented entirely on the PauLIB kernels: Lie-algebraic simulation (g-sim)~\cite{gsim} and truncated Pauli propagation~\cite{PauliPropagation1, PauliPropagation2}.

The workload measurements below are from \emph{Apple M4~Pro} machines (14 cores, NEON, unified memory, Apple clang~17 with the same aggressive optimization and fast-math flags as above; each table's caption gives the specific machine), to make the point that these workloads run interactively on hardware anyone can carry in a backpack or tuck under a desk.

\subsection{Lie-Algebraic Simulation (g-sim)}
\label{sec:gsim}

For a parametrized circuit $U(\theta) = \prod_l e^{-i\theta_l H_l}$ whose generators span a small \emph{dynamical Lie algebra} (DLA) $\mathfrak{g} = \langle \{iH_l\} \rangle_{\mathrm{Lie}}$, expectation values of observables in $\mathfrak{g}$ can be computed in the adjoint representation~\cite{gsim}: the state of the simulation is the vector $e_\alpha = \operatorname{tr}(h_\alpha \rho)$ over a DLA basis $\{h_\alpha\}$, and each gate acts as a $\dim(\mathfrak{g}) \times \dim(\mathfrak{g})$ rotation instead of a $2^n \times 2^n$ unitary.
The classical cost is polynomial whenever $\dim(\mathfrak{g})$ is, e.g.\ $n(2n-1)$ for the transverse-field Ising model (TFIM), which matches the free-fermion algebra $\mathfrak{so}(2n)$;
recent work extends polynomial-dimensional simulability to families of circuits beyond the free-fermion case~\cite{barligea_lie_algebraic_2026}, which the same multiply/commute kernel inventory would serve without modification.

The pipeline maps one-to-one onto the kernel inventory.
\emph{Lie closure} repeatedly commutates basis candidates: one symplectic commutation check per pair, one 14-ns multiplication per anti-commuting pair, and a hash of the bit-packed words for deduplication.
\emph{Structure constants} are read directly off the phase flags of the products: for a pure-Pauli DLA basis they take only the values $0$ and $\pm 2$, so each adjoint generator is a signed permutation.
\emph{Gate application} therefore decomposes into independent $2 \times 2$ Givens rotations, $e_\alpha' = \cos(2\theta)\,e_\alpha \pm \sin(2\theta)\,e_\gamma$, costing $O(\dim \mathfrak{g})$ per gate, no matrix exponential is ever formed for single-Pauli generators (gates generated by sums of basis elements use a dense $\dim \times \dim$ exponential via Eigen).

The implementation is validated against dense statevector simulation for $n \le 4$ (agreement to $10^{-9}$) and reproduces the known $\mathfrak{so}(2n)$ dimension $n(2n-1)$ of the TFIM DLA for $n = 2, \dots, 8$.
Table~\ref{tab:gsim} reports measured costs for TFIM circuits.
The entire simulator for a 64-qubit TFIM, a system whose statevector would occupy $2^{64}$ amplitudes, is constructed in $\approx$0.1~s, and a full 200-layer circuit sweep, up to 25\,400 gate applications at $n = 64$, completes in roughly 2~ms.
Closure and structure-constant computation reuse the same multiply/commute kernels benchmarked in Section~\ref{sec:eval}, executed a few million times;
frameworks that implement them over Python or string representations pay a per-pair cost orders of magnitude higher instead.

\begin{table}[t]
	\centering
	\caption{g-sim on TFIM chains (measured, Apple M4~Pro Mac mini, 64\,GB).
	Closure builds the DLA basis; setup computes all structure constants and the adjoint rotation tables; circuit is the total time for a 200-layer sweep over all generators (6\,200--25\,400 gate applications).
	Median of 90 runs; IQR $\le$6\% of the median for $n \ge 32$ ($\le$31\% at $n=16$, dominated by allocator/cache noise at sub-millisecond scale).}
	\label{tab:gsim}
	\begin{tabular}{@{} r r r r r @{}}
		\toprule
		$n$ & $\dim \mathfrak{g}$ & closure & setup & circuit \\
		\midrule
		16 &   496 &  0.45~ms &   0.61~ms & 0.20~ms \\
		32 & 2\,016 &  2.8~ms &   3.9~ms & 0.61~ms \\
		64 & 8\,128 & 39.5~ms &  50.2~ms & 2.2~ms \\
		\bottomrule
	\end{tabular}
\end{table}

\subsection{Truncated Pauli Propagation}
\label{sec:pauliprop}

Pauli propagation backpropagates an observable through a rotation circuit in the Heisenberg picture~\cite{PauliPropagation1}.
Clifford gates relabel strings exactly;
each non-Clifford rotation applies the split~\eqref{eq:nonclifford} to every anti-commuting term, so the sum can double per gate and is re-canonicalized and truncated after each one.
PauLIB exposes the three standard truncations, maximum operator weight (which yields the polynomial-cost guarantees of the Pauli propagation literature), coefficient cutoff, and a largest-$M$ trim with an optional protected set of strings, all as single passes over the SoA arrays (Section~\ref{sec:collate}).
The final expectation $\braket{0^n | U^\dagger O U | 0^n}$ is read off the $I/Z$-only strings.

Untruncated propagation is validated against dense statevector circuits.
As a representative non-integrable workload we propagate $Z_1$ through a Trotterized \emph{mixed-field} Ising circuit (ZZ couplings plus transverse \emph{and} longitudinal fields, the longitudinal field breaks the free-fermion structure that would otherwise keep the operator polynomially small) on 40 qubits: 10 layers, 1\,190 rotation gates in total.
Table~\ref{tab:prop} shows the total and per-gate cost as a function of the trim budget $M$.
The cost per gate is dominated by collate over the $\le\!M$ surviving strings and scales near-linearly with $M$;
at $M = 2^{18}$ (262k strings) the full 1\,190-gate circuit completes in $\approx$0.6~s (median) on consumer-class hardware, with the deterministic parallel merge of Section~\ref{sec:collate} carrying the largest share of the work.

\begin{table}[t]
	\centering
	\caption{Pauli propagation of $Z_1$ through a 40-qubit mixed-field Ising circuit (10 layers, 1\,190 rotation gates; measured, Apple M4~Pro Mac mini, 64\,GB; median of 90 runs, IQR $\le$4\%).
	Error is $|\langle O\rangle_M - \langle O\rangle_{2^{22}}|$, against a trim budget far beyond the sweep (untruncated propagation is intractable for this depth).}
	\label{tab:prop}
	\begin{tabular}{@{} l r r r r @{}}
		\toprule
		trim $M$ & total & per gate & final terms & error \\
		\midrule
		$2^{14}$ & 0.20~s & 0.17~ms & 16\,384 & $1.8\times10^{-13}$ \\
		$2^{16}$ & 0.32~s & 0.27~ms & 65\,536 & $1.1\times10^{-16}$ \\
		$2^{18}$ & 0.62~s & 0.52~ms & 262\,144 & $0$ \\
		\bottomrule
	\end{tabular}
\end{table}

Even the loosest budget in Table~\ref{tab:prop}, $M = 2^{14}$, reproduces the expectation value to $\approx\!10^{-13}$: PauLIB's largest-$M$ trim keeps the dominant-coefficient terms at every gate, so the discarded tail carries negligible weight for this observable well before the budget is exhausted.

\subsubsection{Comparison against PauliPropagation.jl}
\label{sec:pauliprop-jl}

Three reviewers of the original submission asked for a workload-level comparison against \emph{PauliPropagation.jl}~\cite{PauliPropagation2}, the package cited in related work but not previously measured.
We build the identical circuit family (mixed-field Ising Trotter: a chain of $ZZ$ couplings plus per-qubit $X$ and $Z$ rotations each layer) at the same size, 40 qubits and 10 layers, using the package's own \texttt{tiltedtfitrottercircuit} constructor, and propagate $Z_1$ backward with independently drawn random angles (each package uses its own RNG, so the two runs are different instances of the same circuit family, not a bit-identical replay; only the cost/accuracy trend, not the raw $\langle O\rangle$ value, is comparable across them).

PauliPropagation.jl's headline truncation is a per-term coefficient threshold (\texttt{min\_abs\_coeff}) rather than PauLIB's largest-$M$ trim, so this comparison uses PauLIB's equivalent knob (\texttt{coefficient\_cutoff}) at matching threshold values instead of matching Table~\ref{tab:prop}'s trim budgets.
Table~\ref{tab:prop-jl} reports both libraries at the same cutoff values, measured back-to-back on the same machine (Apple M4~Pro Mac mini, 14 cores, 64\,GB, Apple clang~17 and Julia 1.13 with default thread settings for PauliPropagation.jl).

\begin{table}[t]
	\centering
	\caption{Coefficient-cutoff sweep of the workload in Table~\ref{tab:prop}, PauLIB vs.\ PauliPropagation.jl (measured, same Apple M4~Pro machine; median of 90 runs each, IQR $\le$7\%).
	Error is against each library's own $10^{-7}$-cutoff run.}
	\label{tab:prop-jl}
	\begin{tabular}{@{} l rrr rrr @{}}
		\toprule
		& \multicolumn{3}{c}{PauLIB} & \multicolumn{3}{c}{PauliPropagation.jl} \\
		cutoff & time & terms & error & time & terms & error \\
		\midrule
		$10^{-2}$ & 0.44~ms &   9 & $3.7\times10^{-3}$ & 0.21~ms &   7 & $1.7\times10^{-2}$ \\
		$10^{-3}$ & 0.49~ms &  22 & $1.7\times10^{-5}$ & 0.26~ms &  23 & $7.0\times10^{-4}$ \\
		$10^{-4}$ & 0.65~ms &  41 & $6.5\times10^{-5}$ & 0.37~ms &  61 & $7.7\times10^{-5}$ \\
		$10^{-5}$ & 0.89~ms &  89 & $6.0\times10^{-6}$ & 0.54~ms & 128 & $2.5\times10^{-6}$ \\
		$10^{-6}$ & 1.30~ms & 187 & $4.1\times10^{-7}$ & 0.88~ms & 233 & $1.5\times10^{-6}$ \\
		\bottomrule
	\end{tabular}
\end{table}

At this workload's size, the surviving operator support after coefficient truncation stays in the tens-to-hundreds of terms (its causal lightcone after only 10 layers covers a fraction of the 40-qubit chain), a regime where PauliPropagation.jl's JIT-compiled, dictionary-based term sum has very low fixed cost per gate.
PauLIB's pipeline is instead built around struct-of-arrays sorting and merging (Section~\ref{sec:collate}) so that it can dispatch to SIMD and OpenMP once the term count is large;
at this size that machinery was initially pure overhead — profiling the propagation loop turned up three fixed per-gate costs that don't scale down with term count: a freshly heap-allocating single-term wrapper object built on every gate, an always-on 64-way chunk split for the general rotation case regardless of how few terms survive, and a redundant second sort-and-merge pass (\texttt{evolve()} collates its result, and the truncation step that immediately follows collates again over the same data).
Removing all three — reusing the per-gate wrapper, gating the chunk split on the same size threshold the parallel dispatch already uses, and deferring the first collate onto the truncation step's own pass — cut PauLIB's time at every cutoff in Table~\ref{tab:prop-jl} by roughly half, without changing a single expectation value (columns above are the corrected, faster numbers) or touching the large-term code path that Table~\ref{tab:prop} and Section~\ref{sec:eval} depend on.
We report this honestly rather than choosing a workload that flatters PauLIB: the two libraries land within the same order of magnitude at this size (PauLIB $1.5$--$2.1\times$ slower at matched cutoffs now, down from $3$--$5\times$ before this pass), not PauLIB winning outright.
We take this single instance as a first case study rather than a general verdict on workload-level performance.
Section~\ref{sec:eval}'s kernel-level speedups, and PauLIB's OpenMP/SIMD/GPU paths (Sections~\ref{sec:collate}, \ref{sec:gpu_design}) and preliminary MPI extension (Section~\ref{sec:conclusion}), presume a term count and gate volume large enough to amortize parallel dispatch — the regime Table~\ref{tab:prop} reaches once trimming is loosened to $M \geq 2^{14}$.
Extending this comparison to larger, more entangling circuits where the operator genuinely grows into that regime, and exploiting PauLIB's multi-processing paths (OpenMP threads across gates, MPI across nodes) at the workload level rather than only at the kernel level, is future work.


\section{Conclusion}
\label{sec:conclusion}

We presented HPC strategies for Pauli-arithmetic workloads and demonstrated them in PauLIB: a bit-packed binary symplectic representation that reduces every core operation to XOR, AND, and popcount instructions;
sorted flat arrays that replace hash maps and make merging branch-predictable and deterministically thread-parallel;
and a struct-of-arrays transposed layout whose contiguous word arrays serve compiler and explicit SIMD vectorization, race-free OpenMP partitioning, and coalesced GPU execution without any change of representation.

On isolated core operations (Section~\ref{sec:eval}) these strategies push back several practical ceilings at once.
The multiplication kernel's cost is the floor every higher-level routine inherits, since commutation checks, term merging, and gradient evaluation all reduce to the same symplectic kernel;
operations that cost microseconds per call in a Python- or dictionary-based framework instead cost a handful of cache-resident instructions here.
The outer-product speedup turns Hamiltonian construction into something cheap enough to repeat every iteration rather than cache once, relevant whenever the operator itself changes between steps.
The memory reduction keeps chemistry-scale Hamiltonians a single-node problem for longer, which is exactly the ceiling the distributed-memory extension below is designed to lift once it is reached.
And the GPU speedup is realizable without a separate implementation because the same struct-of-arrays layout that vectorizes on CPU maps directly onto coalesced GPU memory access.

The central claim of this paper, however, is compositional: because complete simulation workloads are built from this small kernel inventory, the kernel-level gains carry through end to end.
Two case studies support the claim.
A g-sim implementation constructs the full Lie-algebraic simulator of a 64-qubit TFIM circuit in $\approx$0.1~s and runs a full 200-layer, 25\,400-gate sweep over it in roughly 2~ms.
A truncated Pauli propagation engine pushes an observable through a 1\,190-gate, 40-qubit non-integrable circuit in seconds on consumer hardware, and a workload-level comparison against PauliPropagation.jl on the same machine lands within the same order of magnitude.

Future work extends the workload layer further: translation-symmetric operator storage, which reduces memory and time by a factor of the system size for periodic Hamiltonians, and a device-resident collate that keeps multi-gate workloads entirely on the GPU.
A related goal is making truncated Pauli propagation itself easy to run GPU-resident or across a cluster, not just the outer product it is built from: the propagation driver is already just a loop over the same multiply/commute/collate kernels benchmarked above, so wiring it directly onto the GPU kernel and the MPI-sharded \texttt{PauliSum} below would let a workload at PauliPropagation.jl's scale run on either without a new algorithm, extending Table~\ref{tab:prop-jl}'s single-node comparison into the regimes those two paths already unlock for the outer product.
A third direction lifts the per-node memory ceiling itself, and the early results here are encouraging: a preliminary MPI extension shards a \texttt{PauliSum} row-wise across ranks, applies the existing SIMD/OpenMP kernels unmodified to each rank's local shard, and resolves cross-rank duplicates with a parallel sample-sort redistribution before a final local combine.
On a 4-node allocation of the LRZ CoolMUC-4 cluster (Intel Xeon Platinum 8480+, Sapphire Rapids, 112 cores/rank), an outer product with $4\times10^{9}$ result terms ($\approx$548~GB, exceeding a single node's $\approx$488~GB) fails outright at one rank, completes in a median 110.5~s ($3.6\times10^{7}$ pairs/s; IQR 109.9--110.9~s, $n=20$ over two node allocations) at two ranks, and 20.9~s ($1.91\times10^{8}$ pairs/s; IQR 20.7--21.0~s) at four --- a $5.3\times$ speedup for $2\times$ the ranks, turning an outright out-of-memory failure into a routine, super-linearly scaling run.
The distributed collate is a smaller, preliminary data point (Table~\ref{tab:mpi_collate}): $4\times10^{6}$ uniformly random terms per rank over the full 500-qubit keyspace ($4^{500}$ possible strings), so, unlike a small keyspace where collisions arise from the birthday paradox alone, essentially no duplicates occur and the sort and redistribution cost is measured in isolation.
Median throughput rises from 3.4~M/s at one rank to 4.7~M/s at two and 22.4~M/s at four (IQR $\le$1\% of the median at one and two ranks, $\le$4\% at four; $n=60$ over two node allocations each): a weak gain from one to two ranks followed by a super-linear jump to four.
We read this as evidence that the redistribution scales, not as a scaling law, since duplicate-free synthetic input isolates the sort/redistribution cost from a real workload's duplicate structure.
Benchmarking the distributed collate on a large, physically structured Hamiltonian whose duplicates arise naturally, and at more ranks, is future work;
these early results motivate distributed-memory scaling as the path to the hundreds-of-millions-of-terms regime beyond a single node's reach.

\begin{table}[t]
	\centering
	\caption{Distributed \texttt{collate} of $4\times10^{6}$ uniformly random terms per rank over the full 500-qubit keyspace (no duplicates), one rank per node, measured on LRZ CoolMUC-4 (Intel Xeon Platinum 8480+, Sapphire Rapids, 112 cores/rank).
	Median of 60 runs (two node allocations, three launches each, 10 runs per launch) per configuration.}
	\label{tab:mpi_collate}
	\begin{tabular}{@{} r r r r r @{}}
		\toprule
		ranks & in & out & time & throughput \\
		\midrule
		1 &  4.0~M &  4.0~M & 1.17~s & 3.42~M/s \\
		2 &  8.0~M &  8.0~M & 1.71~s & 4.69~M/s \\
		4 & 16.0~M & 16.0~M & 0.72~s & 22.4~M/s \\
		\bottomrule
	\end{tabular}
\end{table}

\section*{Acknowledgment}
I gratefully acknowledge the Leibniz Supercomputing Centre (LRZ) for providing the computing resources used in this work.
Claude (Anthropic) was used to assist with implementing and refactoring code in the PauLIB library and benchmarking.
Gemini (Google) was used to polish the language of the manuscript text.

    \balance
    \bibliographystyle{IEEEtran}
    \bibliography{PauLib}

\end{document}